\documentclass[aps,prd,showpacs,floatfix,superscriptaddress,preprintnumbers,nofootinbib]{article}

\usepackage{graphicx,amssymb,url,subfig}

\def\gcmm3{{\,{\rm g\,cm^{-3}}}}

\def\lsim{\raise0.3ex\hbox{$\;<$\kern-0.75em\raise-1.1ex\hbox{$\sim\;$}}}
\def\gsim{\raise0.3ex\hbox{$\;>$\kern-0.75em\raise-1.1ex\hbox{$\sim\;$}}}

\begin{document}

\begin{center}
\vglue .06in
{\Large \bf {Simulations of Ultra High Energy Cosmic Rays propagation}}
\bigskip
\\{\bf O.E. Kalashev \footnote{e-mail: {\tt kalashev@inr.ac.ru}}},
 \\[.05in]
{\it{Institute for Nuclear Research of the Russian Academy of Sciences\\
Moscow, Russia}}
\\[.05in]
{\bf E. Kido \footnote{e-mail: {\tt ekido@icrr.u-tokyo.ac.jp}}},
\\[.05in]
{\it{Institute for Cosmic Ray Research University of Tokyo, Kashiwa, Chiba, Japan}}
\\[.40in]
\end{center}

%\pub{Received (Day Month Year)}{Revised (Day Month Year)}

\begin{abstract}

We compare two techniques for simulation of Ultra High Energy Cosmic Rays (UHECR) propagation, the Monte Carlo approach and method based on solving transport equations in one dimension. For the former we adopt publicly available tool CRPropa and for the later we use the code TransportCR developed by O.K. et al. and used in number of applications which is made available online with publishing this paper. While the CRPropa code is more universal, transport equation solver  has advantage of roughly 100 times higher calculation speed. We conclude that the methods give practically identical results for the case of proton or neutron primaries provided that some accuracy improvements are introduced to CRPropa code.

\end{abstract}

\section{Introduction}
Identification of origin of ultra-high energy cosmic rays is one of the main problems of modern astrophysics.
Although existence of particles with energy $E\gsim10^{19}$ eV has been confirmed by several experiments, their possible sources, propagation mechanism and even their nature are still subjects of intense research. Noticeable progress has been achieved during last decade by the new generation experiments. A suppression of cosmic ray flux  above $E \sim 4 \cdot 10^{19}$\,eV has been observed by HiRes, Telescope Array~\cite{AbuZayyad:2012ru} and Pierre Auger Observatories \cite{AugerCutoff,HiResCutoff} which depending on assumed UHECR composition may indicate either observation of the GZK-effect~\cite{Greisen1964,ZatsepinKuzmin1966}, or natural cut-off in the energy of cosmic ray sources. 
The measurements of the position of the shower maximum and its fluctuations by the Pierre Auger experiment suggest a significant fraction of heavy nucleai above $10^{19}$\,eV~\cite{Abraham:2010yv}
However both composition and energy spectrum studies by HiRes~\cite{HiResCompos} and Telescope Array~\cite{Array:2013dra}) show consistency with pure proton or light element composition in the same energy range.

Ultra-high energy protons and nuclei can not be kept by galactic magnetic field and therefore freely escape galaxy. Since currently there are no known sources within Milky Way which could possibly accelerate protons or nuclei up to ultra-high energies it is assumed that the particles should have extragalactic origin. During their propagation through intergalactic space UHECRs rapidly loose energy in interactions with intergalactic photon background. Understanding the UHECR attenuation process is crucial for making model predictions and interpretation of experimental results.
A number of numerical codes has been developed so far for simulation of UHECR attenuation process and calculation of secondary particle spectra~\cite{CRPropa,propag,Berezinsky:2002nc,Aloisio:2012wj,Allard:2005ha}. The mutual checks
between different tests show consistency level of about 10\%. Due to growing experimental statistics improving simulation accuracy becomes critical.

In this paper we focus on predictions of the spectra from proton sources. We describe in detail our transport equation solving code which has already been used in a number of works~\cite{propag} although hasn't been publicly available until now. We compare our tool with the actively developed Monte Carlo code CRPropa~\cite{CRPropa}, which is publicly available and used in Pierre Auger Collaboration analysis~\cite{ThePierreAuger:2013eja}.
While the former code benefits exceedingly high calculation speed the later is more universal and easy customizable.
In section~\ref{methods} we discuss simulation techniques on which the codes are based. In section~\ref{compare_and_improve} we compare the results of proton propagation simulations and suggest improvements to CRPropa. Finally we make a conclusion.

\section{Calculation of observable spectra}\label{methods}

The observable UHECR spectra calculations for given production scenarios involve simulation of sources and attenuation effects such as interactions of cosmic rays with intergalactic media and their deflection by galactic and intergalactic magnetic fields. The interaction rate  calculation accuracy depends on our knowledge of the infrared intergalactic photon background and it's evolution while particle trajectory calculations rely on the models of intergalactic and galactic magnetic fields. Neither of the above factors is currently known enough to make definitive predictions. On the other hand identification of UHECR sources would help to constrain the properties of intragalactic media which especially applies to magnetic field estimates. Our present knowledge of the intergalactic magnetic field (IGMF) is very poor. The theoretical and observational constraints on the mean IGMF strength $B$ and correlation length $L_{cor}$ are summarised in the review~\cite{Durrer:2013pga}:
\begin{eqnarray}
10^{-17}G \lsim & B & \lsim 10^{-9}G \label{limitB} \\
L_{cor} & \gsim & 1pc  \label{limitLcor}
\end{eqnarray}
The simulation assuming the magnetic field grow in a magnetohydrodynamical ampliﬁcation process driven by structure formation out of a magnetic seed field present at high redshift~\cite{Dolag:2004kp} suggests present IGMF strength $B\lsim10^{-12}G$ (see although Ref.~\cite{Sigl:2003cc}).
One can show (see i.e. Ref.~\cite{Berezinsky:2007kz}) that the effect of magnetic fields on the average energy spectrum of protons with energies $E\gsim10^{18}eV$ is negligible if IGMF strength $B\lsim10^{-10}G$ (assuming $L_{cor}\lsim1Mpc$) provided that the average distance between the UHECR sources doesn't exceed GZK radius. If these conditions are realized the computation of the averaged fluxes can be done by solving the coupled Boltzmann equations for UHECR transport in one spatial dimension or using one-dimensional Monte Carlo simulation (the first method is usually much faster). In the limit of strong magnetic fields when it is important to follow particle trajectories i.e. for calculation of images of discrete sources of UHECR only full 3D Monte Carlo simulation can give reliable result. Our code~\cite{propag} uses formalism of transport equations while the CRPropa~\cite{CRPropa} implements either 1D or 3D Monte Carlo simulations. Below we describe CRPropa code only briefly and pay more attention to transport equation solution.

\subsection{CRPropa}
CRPropa~\cite{CRPropa} is a Monte Carlo simulation tool aimed at studying the propagation of neutrons, protons and nuclei in the intergalactic medium. It provides a one-dimensional (1D) and a three-dimensional (3D) modes. In 3D mode, magnetic field and source distributions can be defined on a 3D grid. This allows one to perform simulations in the source scenarios with a highly structured magnetic field configuration. In 1D mode, magnetic fields can be specified as a function of the distance to the observer, but their effects are restricted to energy losses of electrons and positrons due to synchrotron radiation within electromagnetic cascades. Furthermore in 1D mode it is possible to specify the cosmological and the source evolution as well as the redshift scaling of the background light intensity. All important interactions with the cosmic infrared (IRB) and microwave (CMB) background light are included, namely, production of electron-positron pairs, photopion production and neutron decay. Additionally, CRPropa allows for tracking and propagating secondary $\gamma-$rays, $e^{+}e^{-}$ pairs and neutrinos. The code also contains the module solving one-dimensional transport equations for electromagnetic cascades that are initiated by electrons, positrons or photons taking into account pair production and inverse Compton scattering and synchrotron radiation of electrons. For more details on the code refer to the Ref.~\cite{CRPropa}

\subsection{Transport code}

The code~\cite{propag}  simulates attenuation of protons, neutrons, nuclei, photons and stable leptons by solving transport equation in one dimension taking into account all standard dominant processes. UHE particles lose their energy in interactions with the electromagnetic background, which consists of CMB, IRB and radio components (the last one only effects EM cascade development at ultra-high energies). For IRB backgrounds several models are implemented~\cite{Kneiske:2003tx,Franceschini:2008tp,Stecker:2005qs,Kneiske:2010pt,Stecker:2012ta,Inoue:2012bk}.
For highest energy protons, neutrons and nuclei the main attenuation process is photopion production. Below the threshold of photopion production photodisintegration (for nuclei only) and $e^+e^-$ pair production provide the attenuation mechanism. Although nuclei attenuation is implemented in the code (we use photodisintegration rates derived in~\cite{Stecker:1998ib}), since deflections in magnetic fields can not be precisely described within 1D transport equation formalism, the reliable description of heavy nuclei propagation can be achieved only for energies $E>10^{19}eV$ (assuming $B\lsim10^{-10}G$).  Below we focus on proton and neutron propagation simulations. With the photopion production by protons and neutrons, $e^+e^-$ pair production by protons on background photons and neutron decay included, the transport equations for protons and nucleons can be written in the following form (here and below we assume $\hbar=c=1$):
\begin{eqnarray}
&&\partial_{t}N_p(E_p) = - N_p(E_p) \int
d\epsilon\,n(\epsilon) \int d\mu \frac{1-\beta_p \mu}{2} (\sigma_{\rm
p,\pi}+\sigma_{\rm p,e}) + \nonumber \\
&&\int dE^\prime_{p}N_p(E^\prime_{p}) \int d\epsilon\,n(\epsilon) \int d\mu
\frac{1-\beta^\prime_{p} \mu}{2} 
\left( \frac{d\sigma_{\rm p,\pi}}{dE_p} + \frac{d\sigma_{\rm p,e}}{dE_p} \right)  + \nonumber \\
&&\int dE^\prime_{n}N_n(E^\prime_{n}) \int d\epsilon\,n(\epsilon) \int d\mu
\frac{1-\beta^\prime_{n} \mu}{2} 
\frac{d\sigma_{\rm n,\pi}}{dE_p} + 
N_n(E_p)\frac{m_n}{E_p}\tau_n^{-1} + Q_{\rm p}(E_p), \label{kineq_p}
\end{eqnarray}
\begin{eqnarray}
&&\partial_{t}N_n(E_n) = - N_n(E_n) \int
d\epsilon\,n(\epsilon) \int d\mu \frac{1-\beta_n \mu}{2} \sigma_{\rm
n,\pi} + \nonumber \\
&&\int dE^\prime_{p}N_p(E^\prime_{p}) \int d\epsilon\,n(\epsilon) \int d\mu
\frac{1-\beta^\prime_{p} \mu}{2} 
\frac{d\sigma_{\rm p,\pi}}{dE_n}  + \nonumber \\
&&\int dE^\prime_{n}N_n(E^\prime_{n}) \int d\epsilon\,n(\epsilon) \int d\mu
\frac{1-\beta^\prime_{n} \mu}{2} 
\frac{d\sigma_{\rm n,\pi}}{dE_n} - 
N_n(E_n)\frac{m_n}{E_n}\tau_n^{-1} + Q_{\rm n}(E_n), \label{kineq_n}
\end{eqnarray}
where $N_p(E)$ and $N_n(E)$ are densities of protons and neutrons per unit energy. Here isotropic distribution of background photons is assumed with number density $n(\epsilon)$ depending on photon energy 
$\epsilon$ only, $\beta_p$ and $\beta_n$ are particle velocities, $\mu$ is the collision angle cosine and $Q$ denotes external source terms. The terms describing neutron decay are proportional to $\tau_n^{-1}$, inverse neutron lifetime in the rest frame. In the neutron decay term of eq.~(\ref{kineq_p}) we neglect difference in masses of neutron and proton and assume that the recoiling proton momentum is zero in the neutron rest frame. The terms proportional to $\sigma_{\rm p,\pi}$ and $\sigma_{\rm n,\pi}$ describe photopion production by protons and neutrons. Here we take into account that nature of leading nucleon can be changed in the above interaction. Finally, terms proportional to $\sigma_{\rm p,e}$ describe $e^+e^-$ pair production by protons.

To solve the above equations numerically, we bin the energies of the cosmic rays. We divide
each decade of energy into $n_{d}$ equidistant logarithmic bins. Let us designate the central value of 
the $i$-th bin $E_i$ and boundary values $E_{i-1/2}$ and $E_{i+1/2}$. Than we rewrite equations~(\ref{kineq_p},\ref{kineq_n}) in terms of numbers of particles in each bin
\begin{equation}
N_{\rm p(n),i} = \int_{E_{i-1/2}}^{E_{i+1/2}} N_{p(n)}(E)dE
\end{equation}
After replacing the continuous integrals by finite sums we have

\begin{eqnarray}
\frac{d}{dt} N_{\rm p,i} & = &
- N_{\rm p,i} A_{\rm p,i} + \sum_{j\geq i} B_{\rm p \rightarrow p,ji} N_{\rm p,j} + \sum_{j\geq i} B_{\rm n \rightarrow p,ji} N_{\rm n,j} + Q_{p,i},\label{ode_p}\\
\frac{d}{dt} N_{\rm n,i} & = &
- N_{\rm n,i} A_{\rm n,i} + \sum_{j\geq i} B_{\rm p \rightarrow n,ji} N_{\rm p,j} + \sum_{j\geq i} B_{\rm n \rightarrow n,ji} N_{\rm n,j} + Q_{n,i},\label{ode_n}
\end{eqnarray}
where 
\begin{equation}
Q_{p(n),i} = \int_{E_{i-1/2}}^{E_{i+1/2}} Q_{p(n)}(E)dE \label{coef_q}
\end{equation}
coefficients $A_{\rm p(n),i}$ have physical meaning of interaction rates and are given by
\begin{eqnarray}
&&A_{\rm p,i} =  \int
d\epsilon\,n(\epsilon) \int d\mu \frac{1-\beta_{p,i} \mu}{2} \left[\sigma_{\rm
p,\pi}(E_i,\epsilon,\mu)+\sigma_{\rm p,e}(E_i,\epsilon,\mu)\right] \nonumber \\
&&A_{\rm n,i} = \int
d\epsilon\,n(\epsilon) \int d\mu \frac{1-\beta_{n,i} \mu}{2} \sigma_{\rm
n,\pi}(E_i,\epsilon,\mu) + \frac{m_n}{E_{n,i}}\tau_n^{-1}, \label{coef_a}
\end{eqnarray}
and coefficients $B_{\rm x \rightarrow y,ji}$ are given by
\begin{eqnarray}
&&B_{\rm p \rightarrow p,ji} = \int d\epsilon\,n(\epsilon) \int d\mu
\frac{1-\beta_{p,j} \mu}{2} \int_{E_{i-1/2}}^{E_{i+1/2}}dE_p \times \nonumber \\
&& \left[ \frac{d\sigma_{\rm p,\pi}}{dE_p}(E_j,\epsilon,\mu; E_p) + \frac{d\sigma_{\rm p,e}}{dE_p}(E_j,\epsilon,\mu; E_p) \right], \label{coef_b}\\
&&B_{\rm p \rightarrow n,ji} = \int d\epsilon\,n(\epsilon) \int d\mu
\frac{1-\beta_{p,j} \mu}{2} \int_{E_{i-1/2}}^{E_{i+1/2}}dE_n \frac{d\sigma_{\rm p,\pi}}{dE_n}(E_j,\epsilon,\mu; E_n), \nonumber\\
&&B_{\rm n \rightarrow p,ji} = \int d\epsilon\,n(\epsilon) \int d\mu
\frac{1-\beta_{n,j} \mu}{2} \int_{E_{i-1/2}}^{E_{i+1/2}}dE_p \frac{d\sigma_{\rm n,\pi}}{dE_p}(E_j,\epsilon,\mu; E_p)+
\delta_j^i  \frac{m_n}{E_i}\tau_n^{-1}, \nonumber\\
&&B_{\rm n \rightarrow n,ji} = \int d\epsilon\,n(\epsilon) \int d\mu
\frac{1-\beta_{n,j} \mu}{2} \int_{E_{i-1/2}}^{E_{i+1/2}}dE_n \frac{d\sigma_{\rm n,\pi}}{dE_n}(E_j,\epsilon,\mu; E_n).\nonumber
\end{eqnarray}
The system of ordinary differential equations (ODE)~(\ref{ode_p}) and~(\ref{ode_n}) can be solved using standard methods. The TransportCR code utilizes GNU Scientific Library (GSL) which provides a choice of 11 adaptive step ODE integration schemes. In addition to standard GSL schemes the first order implicit scheme is implemented in the code which benefits the observation that the matrixes $B_{\rm x \rightarrow y,ji}$ in equations~(\ref{ode_p}) and~(\ref{ode_n}) are triangular. This makes it possible to speed up ODE solving by reducing number of independent variables. The implicit method has the advantage that the solution converges even for relatively large time step. However, to ensure the desired accuracy, we need to optimize the step size for a given problem by trial and error. Note that coefficients $A_{\rm x,i}$, $B_{\rm x \rightarrow y,ji}$ depend on time because of the redshift dependence of the background concentration. Since in general the step size should be proportional to interaction length $A^{-1}_{\rm p(n),i}$ and the length itself is inversely proportional to concentration of background photons, we make the time step dependent on z:
\begin{equation}
h(z)=h(0)(1+z)^{-3}
\end{equation}
which corresponds to the evolution of CMB photons concentration. There is no need to recalculate coefficients $A_{\rm x,i}$, $B_{\rm x \rightarrow y,ji}$
before each step unless the background is highly variable. In practice we recalculate the coefficients after intervals of time corresponding to redshift change of one log bin:
\begin{equation}
\frac{z_i+1}{z_{i+1}+1}=10^{1/n_{d}} \label{stepZ}
\end{equation}

The same calculation technique is used to obtain the fluxes of secondary particles produced by nucleons, namely electrons, positrons, photons and neutrino. The electron-photon cascade is driven by chain of inverse Compton scattering and $e^+e^-$ pair production by photons while secondary neutrinos propagate practically without attenuation. The direct application of the above scheme may be difficult in the special case of the fast processes with small inelasticity since it would require high density of energy binning and small time steps. The $e^+e^-$ pair production by protons is good example of such process having inelasticity less then $10^{-3}$. Similar problem occurs in the Monte Carlo simulation method.
In both techniques continuous energy loss (CEL) approximation is used to bypass the problem with mean energy loss rate given by equation
\begin{equation}
-\frac{dE}{dt} = \int d\epsilon\,n(\epsilon) \int d\mu \frac{1-\beta \mu}{2} 
\int dE^\prime (E-E^\prime)\frac{d\sigma}{dE^\prime}(E,\epsilon,\mu; E^\prime)
\end{equation}
While in Monte Carlo simulation the CEL implementation is straightforward in transport equation approach simple first order scheme is used to express continuous energy loss in terms of coefficients $A_{\rm x,i}$, $B_{\rm x \rightarrow x,ji}$ in equations~(\ref{ode_p}) and~(\ref{ode_n}):
\begin{eqnarray}
&&A_{\rm x,i}=\frac{1}{E_{\rm i}-E_{\rm i-1}}dE/dt|_{E=E_{\rm i-1/2}}, \nonumber\\
&&B_{\rm x \rightarrow x,ji}=\delta_{j}^{i+1}A_{\rm x,j}
\end{eqnarray}

Equations~(\ref{ode_p}) and~(\ref{ode_n}) don't take into account expansion of the universe. One way to treat it would be to introduce the CEL term with
\begin{equation}
-\frac{dE}{dt}=-\frac{dE}{dz}\frac{dz}{dt}=\frac{E}{1+z}H_0(1+z)\sqrt{\Omega_m(1+z)^3+\Omega_\Lambda}
\end{equation}
along with replacing $Q_{\rm x}(E)$ with comoving source densities $\tilde{Q}_{\rm x}(E,z)=(1+z)^{-3}Q_{\rm x}(E)$ in~(\ref{coef_q}) to take into account the volume increase. As alternative to introduction of extra CEL term one could perform shifting the energy binning each step~(\ref{stepZ}) as it is done in Ref.~\cite{Lee:1996fp}.
Both methods agree in the limit of large $n_d$, the former is more accurate in the presence of other more rapid attenuation channels while the latter is precise in the absence of any interactions.

\subsubsection{Photopion production}
Photopion production is the main attenuation mechanism for protons and neutrons with energies $E \gsim 10{\,{\rm EeV}}$.
The energy threshold for this process is
\begin{equation}
E_{\rm th} = \frac{2m_{\rm N}m_\pi+m^2_\pi}{4\epsilon} \simeq 6.8 
\left(\frac{\epsilon}{{10^{-3}{\rm eV}}} \right)^{-1} {\,{\rm EeV}}
\end{equation}
where $m_{\rm N}$ is the nucleon mass.
This process was extensively studied in Ref.~\cite{Mucke:1999yb} where the SOPHIA event generator was developed for simulation of photopion production including calculation of various channels cross sections and sampling of secondaries. Both CRPropa and transport equation based code are using SOPHIA event generator as black box. The former is calling the event generator directly while the latter provides the auxiliary routine to calculate the propagation coefficients $A_{\rm x,i}$ and $B_{\rm x \rightarrow y,ji}$ using calls of SOPHIA procedures which is described below.
First let us rewrite the contribution from photopion production in equations~(\ref{coef_a}) and~(\ref{coef_b}) in terms of photon energy $\tilde{\epsilon}$ in the nucleon rest frame (NRF)
\begin{equation}
  \tilde{\epsilon} = \frac{\epsilon E_{\rm N}}{m_{\rm N}} (1-\beta\mu) \equiv  \epsilon \gamma (1-\beta\mu)
\end{equation}
where $m_{\rm N}$ and $E_{\rm N}$ are the nucleon mass and energy respectively and $\gamma=E_{\rm N}/m_{\rm N}$. The threshold energy for photopion production
\begin{equation}
  \tilde{\epsilon}_{\rm th} = \frac{m_{\rm\pi}^2+2m_{\rm\pi}m_{\rm N}}{2m_{\rm N}} \simeq 0.15 {\rm GeV} \label{k_th}
\end{equation}
After omitting in~(\ref{coef_a}) and~(\ref{coef_b}) the terms related to pair production and neutron decay in ultrarelativistic limit ($\beta\rightarrow 1$) we have:
\begin{eqnarray}
&&A_{\rm p(n),i} = \frac{1}{2\gamma_i^2} \int_{\tilde{\epsilon}_{\rm th}}^{2\gamma_i\epsilon_{\rm max}} d\tilde{\epsilon} \,I_b(\frac{\tilde{\epsilon}}{2\gamma})\,\tilde{\epsilon}\,\sigma_{\rm p(n)}(\tilde{\epsilon}),\label{A_pi}  \\ 
&&B_{\rm x \rightarrow y,ji} = \frac{1}{2\gamma_i^2} \int_{\tilde{\epsilon}_{\rm th}}^{2\gamma_i\epsilon_{\rm max}} d\tilde{\epsilon}\, I_b(\frac{\tilde{\epsilon}}{2\gamma})\,\tilde{\epsilon} \int_{E_{i-1/2}}^{E_{i+1/2}}dE_y \frac{d\sigma_{\rm x}}{dE_y}(\tilde{\epsilon}, E_j; E_y),\label{B_pi}
\end{eqnarray}
where $\epsilon_{\rm max}$ is maximal photon background energy in lab frame and
\begin{equation}
I_b(\epsilon_{\rm th}) = \int_{\epsilon_{\rm th}}^{\epsilon_{\rm max}} \frac{n(\epsilon)d\epsilon}{\epsilon^2} \label{Ib}
\end{equation}
is integral depending solely on photon background density which can be tabulated. The total photopion production cross sections as functions of photon energy in NRF $\sigma_{\rm p(n)}(\tilde{\epsilon})$ are explicitly implemented in SOPHIA which is enough for $A_{\rm p(n),i}$ coefficients calculation.

To calculate $B_{\rm x \rightarrow y,ji}$ we create logarithmic binning in $\tilde{\epsilon}$ with $n_d$ steps per decade from $\tilde{\epsilon}_{th}$ given by~(\ref{k_th}) and $\tilde{\epsilon}_{\rm max}=2\epsilon_{max} \gamma_{max}$ and for each $\tilde{\epsilon}_i$ in NRF we perform sampling of secondary particles for $10^5-10^6$ times. Let us assume that the nucleon momentum $p_N$ in the lab frame is directed along the $z$ axes. Than since $p_N\gg\epsilon$  the background photon momentum in NRF $\tilde{\epsilon}$ should point to the opposite direction. Let $\tilde{E}^\prime$ and $\tilde{p}_z^\prime$ be the energy and $z$-component of the sampled secondary particle momentum in NRF. Then it's energy in the LAB frame
\begin{equation}
E^\prime = \gamma (\tilde{E}^\prime + \beta \tilde{p}_z^\prime)
\end{equation}
Therefore in ultrarelativistic limit ($\beta\rightarrow 1$) we have
\begin{equation}
r \equiv \frac{E^\prime}{E_N} = \frac{\tilde{E}^\prime + \tilde{p}_z^\prime}{m_N} \label{fraq}
\end{equation}
From Eq.~\ref{fraq} it follows that the distribution $p(r)$ of the random variable $r$ does not depend on primary nucleon energy $E_N$ and may solely depend on $\tilde{\epsilon}$. Therefore for construction of $B_{\rm x \rightarrow y,ji}$ it is enough to build the 2D tables of $p_{\rm x \rightarrow y}(r; \tilde{\epsilon})$ for each pair of primary and secondary particle. If distribution functions $p_{\rm x \rightarrow y}(r; \tilde{\epsilon})$ are normalized on average total number of secondary particles of type $y$ produced by primary particles of type $x$ in collision with photon $\tilde{\epsilon}$ than
\begin{equation}
B_{\rm x \rightarrow y,ji} = \frac{1}{2\gamma_i^2} \int_{\tilde{\epsilon}_{\rm th}}^{2\gamma_i\epsilon_{\rm max}} d\tilde{\epsilon}\, I_b(\frac{\tilde{\epsilon}}{2\gamma})\,\tilde{\epsilon}\,\sigma_{\rm x}(\tilde{\epsilon}) \int_{E_{i-1/2}/E_j}^{E_{i+1/2}/E_j}dr \, p_{\rm x \rightarrow y}(r; \tilde{\epsilon}). \label{B_sophia}
\end{equation}
In practice the routine tabulates fractions of events with $r$ falling to a given range:
\begin{equation}
P_{\rm l,k} = \int_{10^{(-l-1/2)/n_d}}^{10^{(-l+1/2)/n_d}}dr \, p_{\rm x \rightarrow y}(r; \tilde{\epsilon}_k)\equiv N_{\rm l,k}/N_{\rm tot,k}
\end{equation}

\subsubsection{$e^+e^-$ pair production by protons}
$e^+e^-$ pair production is the main energy attenuation mechanism for protons with energies below the
GZK cutoff. The energy threshold for this process is
\begin{equation}
E_{\rm th} = \frac{m_e (m_p + m_e)}{\epsilon} \simeq 0.5 
\left(\frac{\epsilon}{{\,{10^{-3}\rm eV}}} \right)^{-1} {\,{\rm EeV}}.
\end{equation}
As it was mentioned above the process is characterized by low inelasticity and therefore the CEL approximation is used.
The energy loss rate for the process on arbitrary isotropic background was calculated in Ref.~\cite{PPP}, see formulas (3.11)-(3.19).

\section{Comparison of the simulation results}\label{compare_and_improve}
In this section we use simple phenomenological model for source evolution and injection spectrum to compare the simulations described above:
\begin{equation}
\frac{d\Phi}{dt\,dE} \propto E^{-p} (1+z)^{3+m}, \,E<10^{21}eV,\, z<4 \label{source_model}
\end{equation}

\begin{figure}[t]
%\centering
\subfloat[$m=0$, $p=2.69$]{
\includegraphics[angle=0,width=0.32\textwidth]{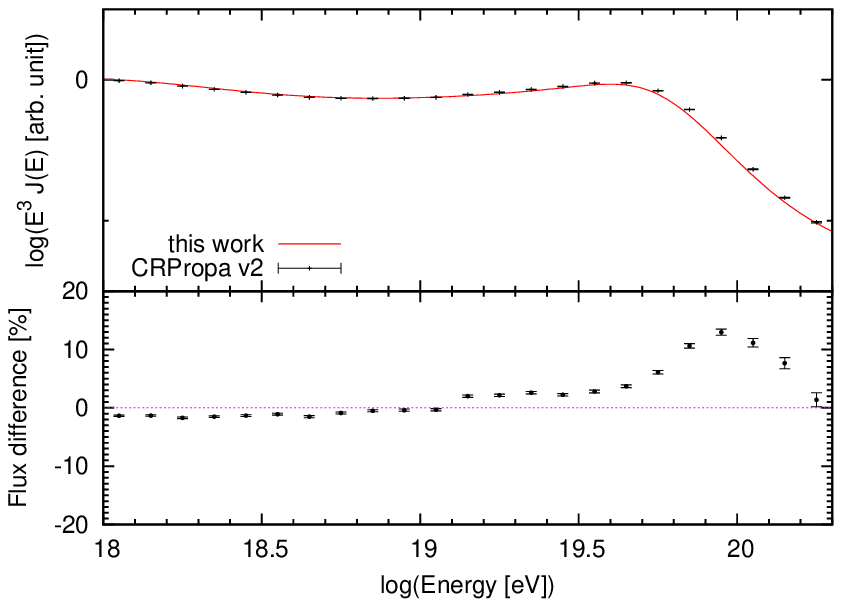}
\label{spec_before_m0}}
\subfloat[$m=4$, $p=2.4$]{
\includegraphics[angle=0,width=0.32\textwidth]{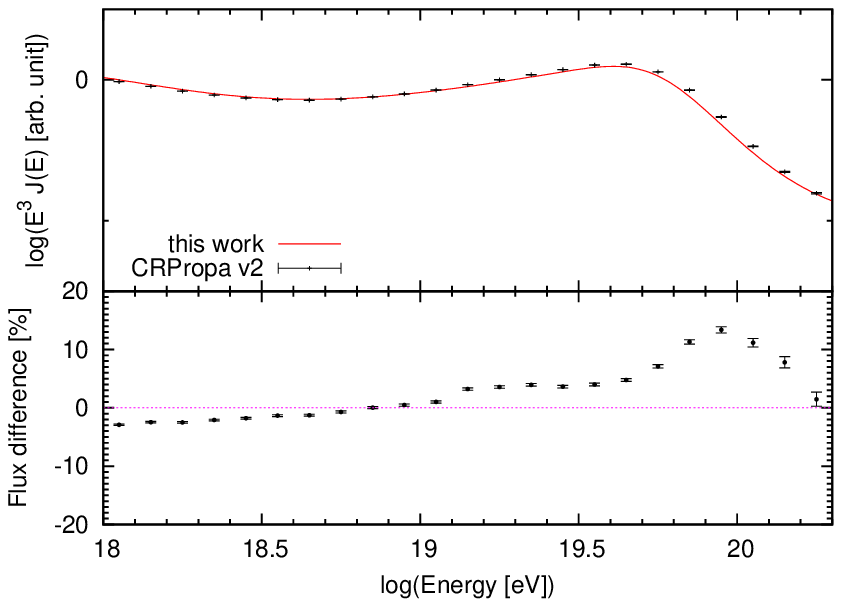}
\label{spec_before_m4}}
\subfloat[improved $m=4$, $p=2.4$]{
\includegraphics[angle=0,width=0.32\textwidth]{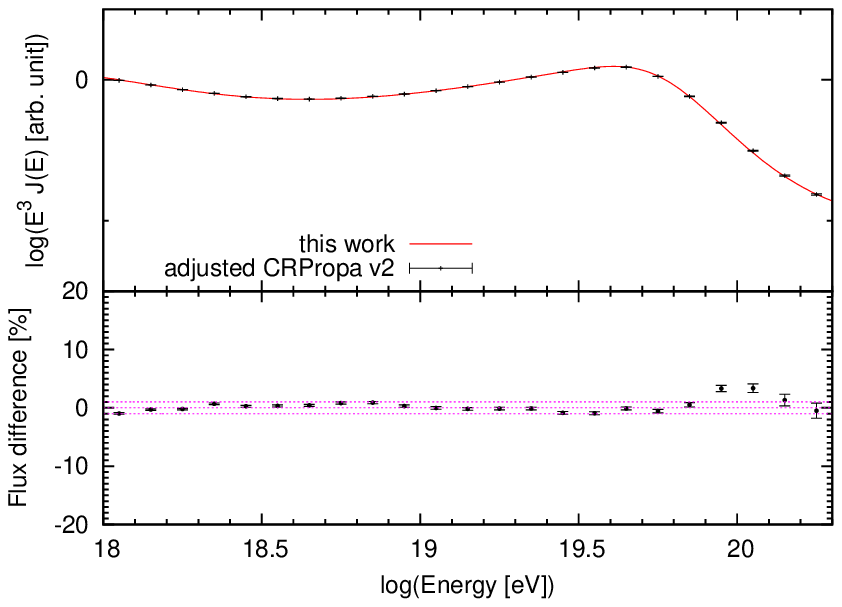}
\label{spec_after}}
\quad
\caption{Propagated spectra calculated by CRPropa v2 (shown in black) and kinetic eq. based code with binning density $n_{d}=100$ (red line) for  the source model parameters listed in captions and assuming EBL model of Ref.~\cite{Kneiske:2003tx}. Lower panels show relative difference in the spectra predictions. Figures $(a,b)$ were obtained using original CRPropa ver.2 while for Fig.$(c)$ the corrected version of  CRPropa was used. Also $1\%$ error band is shown in Fig.$(c)$ in pink.
} \label{spec}
\end{figure}

If Fig.~\ref{spec} we show propagated spectra calculations in two numerical codes. We consider both sources with constant density (Fig.~\ref{spec_before_m0}) and the sources with strong evolution (Fig.~\ref{spec_before_m4}). The highest discrepancy is observed in the later case. It reaches $14\%$ at super-GZK energy. Nevertheless the effect of this discrepancy on the spectrum fitting is weaker than that of uncertainty in sub-GZK energy region where more statistics is available. In this region the difference in the flux predictions is at most $4\%$. In the case of nonevolving source the discrepancy is less pronounced. The above observations naturally lead to assumption that the differences may be related to pion production and in particular the implementation of this process for $z>0$.

\begin{figure}[t]
%\centering
\subfloat[Pair production energy loss rate at $z=1$]{
\includegraphics[angle=0,width=0.5\textwidth]{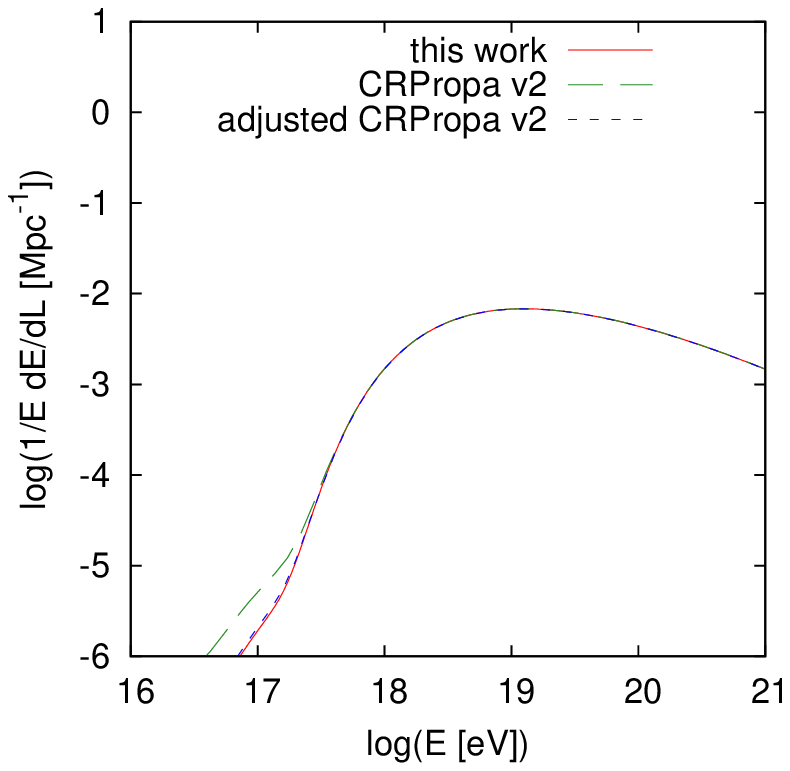}
\label{ppp_old}}
\subfloat[Photopion production rate at $z=1$]{
\includegraphics[angle=0,width=0.5\textwidth]{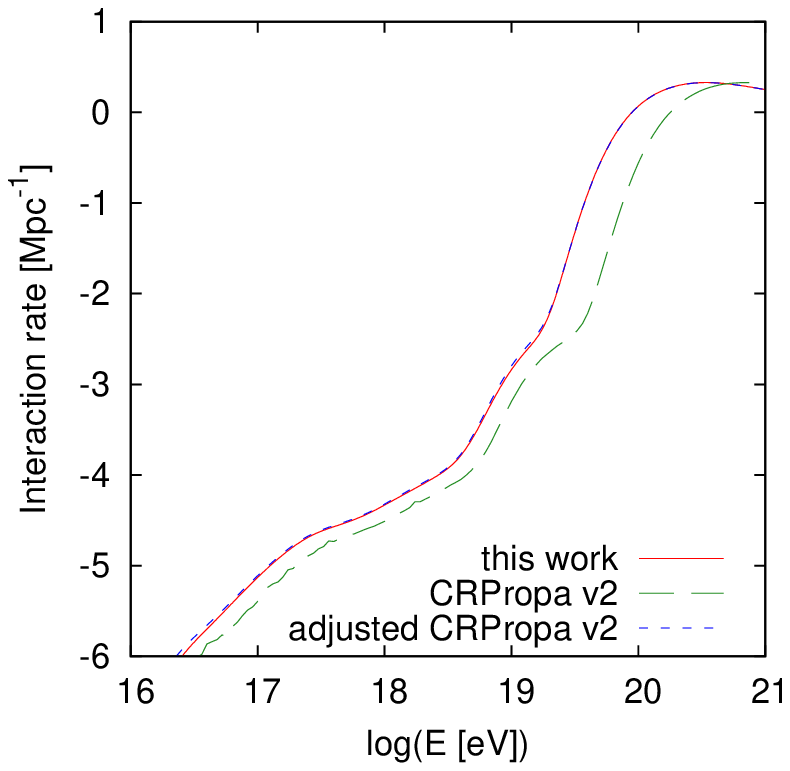}
\label{pi_old}}
\quad
\caption{Comparison of rate calculations between two programs. The energy loss rate for $e^+e^-$pair production process (left figure) and interaction rate for pion production process  (right figure) calculated for $z=1$. Rates obtained in CRPropa v2 are shown in green,  the rates calculated in transport eq. code shown in red and the rates recalculated in CRPropa after corrections described in the text are shown in blue (EBL model of Ref.~\cite{Kneiske:2003tx} in assumed). 
} \label{rates}
\end{figure}

In Fig.~\ref{rates} we compare the energy loss rates for $e^+e^-$~pair production and the interaction lengths for pion production (see Fig.~\ref{rates}) at redshift $z=1$.
The discrepancy is clearly seen for both processes in Figs.~\ref{rates},  although in case of pair production it shouldn't have effect on the spectrum since the energy loss lengths become different in the region where redshift is the main attenuation mechanism. In fact the discrepancy is caused by the simplifying assumption on the dependence of the energy loss rate on $z$ used in CRPropa:
\begin{equation}
\frac{1}{E}\frac{dE}{dt} (E,z) =  (1 + z)^3 \frac{1}{E}\frac{dE}{dt} (E(1+z), z=0) \label{CMBrateEvol}
\end{equation}
which is only valid for CMB background.

The difference in the pion production interaction length is more important. In CRPropa code pion production on CMB is implemented using prebuilt interaction rate tables for $z=0$ and scaling with redshift according to formula similar to~(\ref{CMBrateEvol}), while interaction rates on infrared background are calculated for each $z$ in the same way as pion production term in formula~(\ref{coef_a}) namely by integrating the collision angle averaged cross section (which is tabulated) with photon background spectrum. After inspecting CRPropa code and tables we came to conclusion that the cross section function should be tabulated for broader range of arguments, besides the integration procedure accuracy as well as cross section tables interpolation can be enhanced. With the above corrections implemented we have also rebuilt the interaction rates table for CMB background.
In Fig.~\ref{rates} the corrected rates are shown as well. We've also implemented more precise calculation of pair production energy loss rates in CRPropa which takes into account the evolution of infrared background in proper way.

Fig.~\ref{spec_after} illustrates level of agreement achieved after applying corrections to CRPropa code in case of strong source evolution. Note that in case on  non-evolving source the same level of accuracy is achieved. Calculation of spectra presented in Fig.~\ref{spec} took 60 hours of 2.2 GHz CPU time using CRPropa and 0.6 hours CPU time using Transport code.

We have compared the results of simulating ultra-high energy nucleons propagation using CRPropa and TransportCR codes and suggested improvements to the procedure of interaction rate calculation in CRPropa. After applying the suggested improvements we have achieved $1\%$ level of agreement in flux predictions by the two codes for the whole relevant energy range except small interval in super-GZK region $10^{19.9}\leq E\leq 10^{20.1}$ where relative error grows to $3\%$. The level of accuracy achieved is enough for consistent analysis of the latest UHECR data. We have applied the propagation codes described above to fit the Telescope Array experimental spectrum~\cite{TAfit} assuming phenomenological source model~(\ref{source_model}). The enhancements in CRPropa introduced in this work let us achieve the systematic uncertainty in the best fit parameters related to the choice of propagation code at level $\Delta p\simeq0.01$ and $\Delta m\simeq0.1$. Modified CRPropa and Transport code described in this work can be downloaded from~\cite{sources}. The modifications suggested in the former code have been discussed with CRPropa developers team and are going to be incorporated into CRPropa 3 release version~\cite{guenter}.

%\section{Acknowledgments}
%%%%%%%%%%%%%%%%%%%%%%%%%%%%%%%%%%%%%%%%%%%%%%%%%%%%%%%%%%%%%%%%%%%%%%
{\em Acknowledgments---}%
%%%%%%%%%%%%%%%%%%%%%%%%%%%%%%%%%%%%%%%%%%%%%%%%%%%%%%%%%%%%%%%%%%%%%%
We thank Masaki Fukushima, Kazumasa Kawata, Hiroyuki Sagawa, G{\"u}nter Sigl and Igor Tkachev for fruitful discussion.
The work of O.K. was supported by the Russian Federal Science Fund  grant RSCF 14-12-01340.

%%%%%%%%%%%%%%%%%   figures  
\end{document}